\documentclass[12pt]{article}

\usepackage{graphicx}

\newcommand{\h}{{\eta}}
\newcommand{\s}{{\sigma}}

\newcommand{\cB}{{\cal B}}
\newcommand{\cE}{{\cal E}}
\newcommand{\bN}{{\bf N}}
\newcommand{\bR}{{\bf R}}
\newcommand{\Z}{{\bf Z}}
\newcommand{\sset}{\subset}
\newcommand{\pmu}{\{-1,1\}}

\begin{document}   
\title{
Droplet Motion for the Conservative 2D Ising Lattice Gas
Dynamics below the Critical Temperature}

\author{
Giorgio Favrin$^{(1)}$, Enzo Marinari$^{(2)}$, and 
Fabio Martinelli$^{(3)}$\\[0.5em]
\small $^{(1)}$: Department of Theoretical Physics, Lund University\\
\small S\"olvegatan 14A, S-22362 Lund, Sweden\\
\small $^{(2)}$: Dipartimento di Fisica, INFM and INFN,
      Universit\`a di Roma {\em La Sapienza}  \\
\small P. A. Moro 2, 00185 Roma, Italy\\
\small $^{(3)}$: Dipartimento di Matematica, Universit\`a di Roma 
         Tre\\
\small Largo S. Murialdo 1, 00146 Roma , Italy\\[0.3em]
\small e-mail: {\tt favrin@thep.lu.se, Enzo.Marinari@roma1.infn.it}\\
\small {\tt martin@mat.uniroma3.it}
}


\maketitle
 
\begin{abstract}
\noindent 
We consider the $2D$ Ising lattice gas in a square of side
$L$ with free boundary conditions, temperature below the critical one
and particle density slightly above the density of the vapor
phase. Typical configurations consist of a quarter of a Wulff droplet
of the liquid phase centered at one of the corners of the given
square. We then introduced a reversible Markovian spin exchange
dynamics, also known as Kawasaki dynamics, on the configuration space
and we discuss the heuristics of the transition of a bubble of the liquid
phase from one corner to another. We then present some numerical
evidence suggesting that the preferred mechanism to make the transition
is via evaporation of the original bubble and simultaneous
reconstruction of a new bubble around a new corner.
\end{abstract}

\section{Introduction}

The problem of computing the relaxation time of stochastic Monte Carlo
algorithms for models of lattice classical spin models has attracted
in the last years considerable attention and many new rigorous
techniques have been developed giving rise to nice progresses in
probability theory and statistical mechanics. If for simplicity we
confine ourselves to $\pm 1$ (or $0,1$ in the lattice gas picture)
spins, the two most studied random dynamics have been non-- conservative
Glauber type algorithms, in which a spin at a time flips its value with
a rate satisfying the detailed balance condition w.r.t. the grand
canonical Gibbs measure, and conservative Kawasaki dynamics in which
nearest neighbors spins exchange their values with a rate satisfying the
detailed balance condition w.r.t. the canonical Gibbs measure.
  
A warning sign here is in order.  Most of the rigorous results presented
below have been obtained in the ``continuous time'' setting, namely when
the stochastic dynamics is a continuous time Markov chain and the time
units are such that, in a unitary time interval on average every spin
tries to change its value once. In the physics literature, but also in
the Markov chain Monte Carlo approach to computational problems and image
reconstruction, the discrete time setting in which at each time step
only one dynamical variable is updated is more familiar. In most cases
the simple rule to translate results from one case to another is to
multiply by the appropriate change of scale, but sometimes subtle
problems may appear in the discrete setting (see e.g. \cite{FMS}). 

Let us go back to a quick review of the main results.  For Glauber
dynamics in the one phase region the general picture is relatively clear
for a wide class of models and the conclusion is that equilibrium is
reached exponentially fast provided certain mixing assumption are
satisfied by the grand canonical Gibbs measure (see e.g. \cite{M} and
references therein).  In the conservative case, instead, the basic
results \cite{LY}, \cite{Y} and more recently \cite{CM}, state that, at
high temperature, the relaxation time in a box of side $L$ grows like
$L^{2}$ (diffusive scaling).

A natural question arises as to what happens when the thermodynamic
parameters (e.g inverse temperature and the external magnetic field in
the Glauber case or inverse temperature and particle density in the
conservative case) are such that we do have a phase transition in the
thermodynamic limit.

Let us start with the Glauber case and, to be concrete, let us consider
the usual Ising model in a two dimensional box $V$ of side $L$, without
external field and inverse temperature $\beta$ larger than the critical
value $\beta_c$. With free boundary conditions the picture of the
relaxation behavior to the Gibbs equilibrium measure that comes out is
the following. The system first relaxes rather rapidly to one of the two
phases and then it creates, via a large fluctuation, a thin layer of the
opposite phase along one of the sides of $V$. Such a process
requires an average time of the order of $\exp(\beta\tau_\beta L)$ where
$\tau_\beta$ denotes the surface tension in the direction of one of the
coordinate axes.  After that, the opposite phase invades the whole
system by moving, in a much shorter time scale, the interface to the
side opposite to the initial one and equilibrium is finally reached.
The time required for this final process can be computed to be of the
order of $L^3$ at least in the SOS approximation (see \cite{Po}).

Let us now turn to the much more involved conservative case in the same
setting (low temperature and free boundary conditions). Here
results are much less precise (see
\cite{CCM}).

Let us denote by $N$ the total number of particles in $V$ and let
us assume that $\rho \in (\rho^*_-,\rho^*_+)$, where $\rho$,
$\rho^*_{\pm}$ denote the actual density and the density of the
liquid/vapor phases respectively. With these assumptions it can be
shown that the relaxation time is again exponentially large in $L$ but,
contrary to the conservative case, almost nothing is known rigorously
about the exact constant in the exponential and about the physical
mechanisms behind equilibration. 

That the relaxation time is very large can be easily understood (but
painfully proved) by the following reasoning. Assume for simplicity
that $\rho$ is slightly above $\rho^*_-$ in such a way that,
typically, the gas at equilibrium shows phase segregation between a
(macroscopically) small bubble of liquid immersed in vapor. Because
of the free boundary conditions the bubble prefers to sit around one
of the corners of $V$. Clearly, in order to reach equilibrium, the
system started as above has to move the bubble to the other corners
but, in doing that, it is forced to explore a region of the phase
space of exponentially (in $L$) small canonical probability (and that
is the hard part in a rigorous approach)).  In other words it seems
clear that the slowest mode in the dynamics comes from the corner to
corner droplet transport.  A precise understanding of the above motion
appears therefore to be an interesting problem on its own and almost
unavoidable if one wants to quantify precisely the relaxation time. A
first numerical attempt in this direction represents the main goal of
this note, which is in turn based on the work contained in \cite{F}.

We conclude by observing that, if boundary conditions are changed to e.g
plus or periodic, then the whole scenario changes drastically and one may
argue that the slowest mode of the system is associated to the random
walk motion of the center of gravity of the unique Wulff bubble of the
liquid phase. If that is true then simple heuristics shows that the
relaxation time should now grow as $L^{3}$ (in continuous time units) or
$L^5$ in discrete time steps.

\section{Model and Notation}

We define here our model and notation.

\subsection{Equilibrium Probability Measures}

We consider the two dimensional lattice $\Z^2$ with {\it sites\/} $x =
(x_1, x_2 )$.  By $Q_L$ we denote the square of all $x=(x_1,x_2) \in
\Z^2$ such that $x_i \in \{ 0, \ldots, L-1 \}$, $i=1,2$.  
The {\it edges} of $\Z^2$ are those $e=[x,y]$ with $x,y$ nearest
neighbors in $\Z^2$.  Given $V\sset \Z^2 $, we denote by $\cE_V$ the set
of all edges such that both endpoints are in $V$.

{\it The configuration space.}  Given $V \sset \Z^2$, our {\it
configuration space} is $\Omega_V = S^V$, where $S=\pmu$. Sometimes the
{\it lattice gas\/} point of view will be more convenient, so we also
consider the space $\Omega'_V =
\{0,1\}^V$ and its natural one--to--one correspondence with $\Omega$.
We define the {\it unnormalized magnetization}
$M_V : \Omega_V \mapsto \Z$ and the {\it number of particles\/} $N_V: \Omega_V'
\mapsto \bN$ as

\begin{equation}
  M_V(\s) = \sum_{x\in V} \s(x) \,,
  \qquad
  N_V(\h) = \sum_{x\in V} \h(x)
  \label{mag}
\end{equation}
while the {\it normalized magnetization} is given by $m_V = M_V/ |V|$.

{\it The interaction and the Gibbs measures.}  Given a finite set
$V\sset \Z^2$ we define the Hamiltonian with free boundary condition
$H_V : \Omega_V \mapsto \bR$ by

\begin{equation}
  H_V(\s) = 
  - \sum_{[x,y] \in \cE_V}  \s(x) \s(y) \,.
  \label{H}
\end{equation}
The corresponding (finite volume) Gibbs measure is given
by

\begin{equation}
    \mu^\beta_V(\s) = 
    \bigl(Z^{\beta}_V\bigr)^{-1}
    \exp[ \,-  \beta H_V(\s) \,] \ ,
    \label{finvolmea}
\end{equation}
where $Z^{\beta}_V$ is the proper normalization factor called partition
function.  We denote by $m^*(\beta)$ and $\beta_c$ the {\it
spontaneous magnetization\/} and {\it inverse critical temperature}
of the two dimensional Ising model respectively. It is well known that
$\beta_c = (1/2) \log( 1 + \sqrt{2})$. The density of the liquid and the
vapor phase, denoted respectively by 
$\rho_+^*(\beta)$ and
$\rho_-^*(\beta)$, can be expressed as ${1\over 2}(1\pm m^*)$.

We also introduce the {\it canonical Gibbs measure\/} defined as

\begin{equation}
  \nu^\beta_{V, N}= \mu_V^{\beta}( \cdot \, | \, N_V = N ) \qquad 
  N \in \{ 0, 1, \ldots, |V| \} \ ,
  \label{cano}
\end{equation}
where $N_V$ is the number of particles in $V$.

\section{The Kawasaki Dynamics}

Here we define the relevant Markovian dynamics, reversible with
respect to the canonical Gibbs measure, that will be used and analyzed
in the rest of this paper. 

Given $V \sset \Z^2$, $b\in\cE_V$, and a particle
configuration $\eta\in\Omega'_V$ (equivalent to a spin configuration
$\sigma\in\Omega_V$), let $\eta^b$ be the configuration obtained from
$\eta$ by interchanging the values of the $\eta$ variables at the end
points of the bond $b$. The energy difference between the two
configurations is given by

\begin{equation}
  \Delta_b H_V(\eta) \equiv  H_V(\eta^b) -  H_V(\eta)\ .
  \label{deltaE}
\end{equation}
If $\eta^b\ne\eta$ we call $b$ a {\it broken bond} and we say it
belongs to $\cB_\eta$.  With the above notation the Kawasaki dynamics
with Metropolis transition probability matrix is given by:

\begin{equation}
  W(\eta,\eta') = 
  \cases{
    {1 \over |\cE_V|} e^{-\beta \max(0, \Delta_b H_V(\eta))}
         & if $\eta'=\eta^b$ and $b \in \cB_\eta$\ , \cr
    1-\sum_{b\in\cB_\eta} W(\eta,\eta^b) 
         & if $\eta'=\eta$\ ,\cr
    0  
         & otherwise\ .\cr
  }
  \label{pesoW}
\end{equation}
Clearly, for each $N\in [1,|V|-1]$, $W$ describes an ergodic Markov
chain on the configuration space $\Omega'_{V,N}$ which consists of all
particle configurations with particle number equal to $N$. In
particular, since $W$ satisfies the detailed balance condition with
respect to the canonical measure $\nu_{V,N}^\beta$, the latter
coincides with the unique invariant measure of the chain.

\section{Corner to Corner Matter Transport}

In this section we will first try to define our field of investigation
by starting with a discussion of the typical configurations for the
canonical measure of the Ising model below the critical temperature
with free boundary conditions.  We will then analyze some possible
mechanisms for the corner to corner bubble transition, and the typical
time scales that govern the process.  Subsequently we will discuss the
details of our numerical simulations and give some hints about the
updating algorithm.  Finally we will present the main
results of this note.

\subsection{Heuristics}

Let $V=Q_L$, let $\rho \in (\rho^*_-,\rho^*_+)$, where $\rho^*_{\pm}$
have been defined above, and let $N = \lfloor\rho L^2\rfloor$ be the total
number of particles.

For the above situation, it is useful to recall first the shape of the
typical configurations of the canonical Ising Gibbs measure with free
b.c. when the temperature is below the critical value and $L$ is very
large.

Let $m_\rho = 2\rho -1$ be the usual magnetization associated to the
given particle density. Then, as discussed in \cite{Sh} and
\cite{CGMS}, there exists $0< m_1(\beta) < m^*(\beta)$ such that

\begin{enumerate}
\item{(i)} if $m_\rho \in (-m_1,m_1)$ then the typical configurations 
show phase segregation between a high density ($\approx \rho^*_+$)
region and a low density ($\approx \rho^*_-$) region that are roughly
two horizontal (or vertical) rectangles of appropriate area separated by
an horizontal (or vertical) interface of length $\approx L$.
\item{(ii)} if $m_\rho \in \left(-m^*(\beta),-m_1(\beta)\right] 
\cup \left[m_1(\beta),m^*(\beta)\right)$ 
then the typical configurations show phase segregation between a high
density ($\approx \rho^*_+$) region and a low density ($\approx
\rho^*_-$) region, the smaller of which is a quarter of a Wulff shape
(see \cite{DKS}) of appropriate area and it is centered in one of the
four vertices of $Q_L$.  For the reader convenience we recall that a
Wulff shape is, at very low $T$, very similar to a perfect square.
\end{enumerate}

What is important for us is that in both cases the typical
configurations of the canonical measure show a discrete symmetry
described by rotations of $k{\pi\over 2},\; k=0,1\dots$ around the
center of $\Lambda$.  As a consequence, if the dynamics starts from one
typical configuration for which for example the particles form a
cluster in a corner (this is one of the situations we have described
before), then, in order to reach equilibrium, it must necessarily
cross an unlikely region in the configuration space. One can show that
the canonical probability of such region is exponentially small in $L$
\cite{CCM}. Therefore a bottleneck is present in the configuration
space, and the relaxation time is exponentially large in $L$
\cite{CCM}.

In the sequel of this note we will always assume to be in the second
between the two scenario's described above, and in particular we will
assume that $m$ and $\rho$ are respectively larger but quite close to
the values $-m^*(\beta)$ and $\rho_-^*(\beta)$.  In other words we are
considering a situation in which typically the particles form a small
(on a macroscopic scale) cluster of a specific shape around one of the
four corners of $\Lambda$.  Clearly we do not mean that all the particles
that are in $\Lambda$ are in the bubble, and we do not mean that the bubble
does not have empty holes in its interior. What we mean exactly is that,
with high probability, there exists a macroscopic region with a precise
shape around one of the corners where the particle density is very close
to the density of the liquid phase, $\rho_+^*$ (corresponding to the
Onsager magnetization density).

Next we analyze some possible mechanisms for the corner to
corner bubble transition, namely the process that moves the liquid
bubble from one corner to a different one. 

We can see two main mechanisms that could intervene.

\begin{figure}
\centering
\includegraphics[bb= 230 0 721 171,width=0.4\textwidth,angle=0]{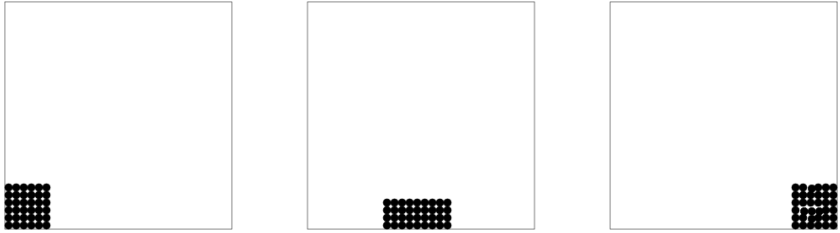}
\caption[a]{
The sliding motion of the bubble, from the left to the right bottom corner.
\protect\label{FIG-1}}
\end{figure}

The first one involves a deformation of the initial Wulff droplet,
that for simplicity we will assume here and in the following to be a
square of side $B$ (where $\rho L^2 = \rho^*_+B^2+\rho^*_-(L^2-B^2)$)
and with one vertex sitting on the left lower corner of $\Lambda$,
into a rectangle. We call this case {\it sliding}.  We can describe it
as follows: we assume that the particles always (that is also during
the transition to one of the other corners) form one compact cluster
(apart from the usual small fluctuations) that, based on energetic
considerations, we can assume to be a generic rectangle $\cal R$ of
sides $B_1$ and $B_2$. Energetically, because of the free boundary
conditions, at equilibrium it will find convenient to be a square of
side $B$ attached to two sides of $\Lambda$. Under the Kawasaki
dynamics we can imagine that such square gets deformed, by some sort
of matter transport along the boundary, until it reaches the shape of
a rectangle $\cal R^*$ of sides $B_1 > B_2$ ($B_1$ being the
horizontal side) with the left and bottom sides attached to the
boundary of $\Lambda$. At this point the rectangle $\cal R^*$ begins
to slide along e.g. the bottom side of $\Lambda$ until it reaches the
opposite corner where it deforms back to the original square (see
figure (\ref{FIG-1})). The energy barrier $\Delta H$ one has to cross
in this process is clearly of the order of $2(B_1 +2B_2) - 2(2B)$ at
least at very low temperature. If we now optimize over $B_1$ and $B_2$
under the constraint that $B_1 \times B_2 = B^2$ we get $B_1=
\sqrt{2}B$ and $B_2={1\over \sqrt{2}}B$.  Thus $\Delta H=
4(\sqrt{2}-1)B$ and therefore we expect that the average time to see a
{\it sliding} transition to be of the order $t_{sliding} \approx
\exp(\beta \Delta H)$.

The second mechanism that one can imagine is what we call {\it
evaporation}. Particles individually separate from the liquid bubble
(they evaporate) and start to perform a sort of random walk in the vapor
phase (typically along the boundaries, by energetic reasons). Clearly
the excess particles in the vapor phase prefer, as soon as they can, to
gather together around one of nearby corners, and grow there a
new liquid bubble. Typically these trial bubbles do not achieve a
macroscopic volume, but die much before reaching this stage and their
particles end up to go back to their mother bubble. Only rare
fluctuations lead to more than half of the particles clustering
together: in these cases typically the new bubble will form in the
selected corner, and evaporation will continue until all the original
liquid bubble has been reformed around the new corner (clearly particles
of the bubble get interchanged at all moments with particles that are in
equilibrium in the vapor phase). It is not difficult to check that, at
least to leading order, the energy barrier is crossed when the two
groups of particles in the two different corners have the same volume
(and same Wulff shape). At very low temperature we get immediately that
$\Delta H= 4 (\sqrt{2} - 1) B$, exactly as in the former case.

We conclude this part by stressing that all the above reasonings
were based only on energy barrier considerations and that we have never
taken into account entropic contributions. The latter may play an
important role in selecting one mechanism instead of another. Moreover
prefactors in the typical time scales of each process may also be
relevant and in that case a more detailed analysis would be required.

\subsection{Numerical Simulations}

Our experimental setting can be described as follows: we use a square
two dimensional lattice with free boundary conditions. We have
investigated lattice sizes going from $L=20$ to $L=30$. The inverse
temperature $\beta$ has been assigned the two values $\beta=0.7$ and
$\beta=1.05$ in different cases (this is of the order of twice the
value of the inverse critical temperature $\beta_c$). Finally we have
chosen the number of particles in the range $25$ to $36$
(corresponding to initial conditions with a bubble of $5\cdot5$ or
$6\cdot 6$ particles in a corner of the square lattice). Notice that
in our range of temperatures the density of the vapor $\rho^*_-$ lies
between $5\ 10^{-3}$  and $2\ 10^{-4}$ for the two different
temperature values: in other words we are in extreme conditions.

The updating algorithm is of course based on the transition matrix
(\ref{pesoW}), and goes as follows.  We consider the particle
configuration at time $t$, $\eta(t)$.  We select a broken bond (that
is a bond in the set ${\cal B}_{\eta(t)}$) at random, with uniform
probability. We compute the energy difference $\Delta_bH_V(\eta(t))$
defined in (\ref{deltaE}): if it is negative we accept the update
proposal, and set $\eta(t+1)= \eta^b(t)$. Otherwise we accept the
update proposal with probability $exp(-\beta\Delta_bH_V(\eta(t)))$ ,
and refuse it otherwise, setting in this case $\eta(t+1)= \eta(t)$.

This procedure, while appealing from the point of view of the
computational efficiency, and satisfactory from the physical point of
view, does not satisfy detailed balance, since the number of broken
bonds is not a conserved quantity (see for instance the discussion in
\cite{ShHe}). In our working conditions, however, this effect is very
small and as far as many issues are concerned, irrelevant. In fact, as
we have already explained, we will mostly focus on the mechanisms on
which the bubble transition is based: since the small violation of
detailed balance amounts to watching a movie that runs at slightly
variable speed, spatial phenomena (like for example which is the
transition path of the bubble or which is its typical spread during
the transition) are described in an exact manner. Scaling arguments
(for example the scaling of the transition time with $\beta$) could as
matter of principle be sensitive to the violation of detailed balance:
we believe however that only the prefactors will be affected, and all
of the many tests of universality that we have done on our simulations
do confirm this point of view.

We still have to discuss the criterium by which we define a bubble
transition. We have indeed used two different criteria and checked
that in the two cases we get consistent results. In the first scheme
we define four boxes, with a vertex in one of the four corners, with
an area equal to the one of the original bubble (that starts, let us
say, from the top left corner). We then define a transition as the
event where the center of mass of the particles that constitute the
bubble enters a new box. In the second scheme we define a transition
as the event where the $3 \over 4$ of the particles have left the
initial box: since we expect to have a transition when half of the
particles have left, we are confident that $75$ percent is a safe
signal for a transition.  Numerical simulations will confirm the
coincidence of the two criteria.

\section{Results and Discussion}

\begin{figure}
\centering
\includegraphics[width=0.7\textwidth,angle=0]{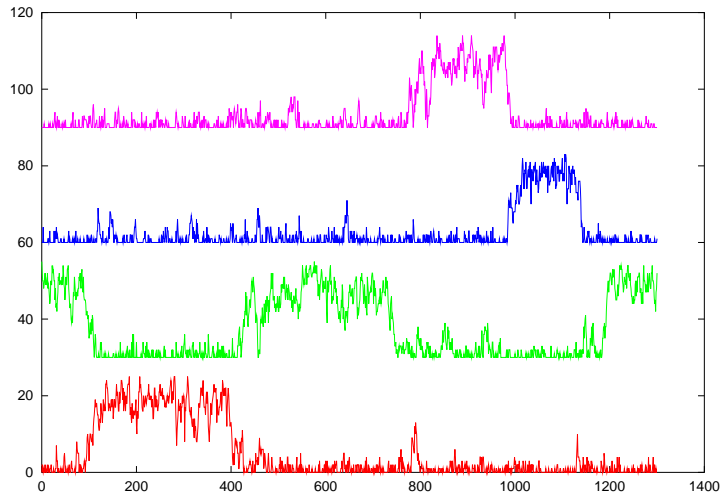}
\caption[a]{
The number of particles close to each of the four corners
as a function of time. See the text for further details.
\protect\label{FIG-2}}
\end{figure}

In figure (\ref{FIG-2}) we show the number of particles (modulo a
corner dependent offset needed for making the figure readable) at time
$t$ near the 4 different corners of the lattice during the course of
the dynamics. More precisely we have defined square boxes of size
$B^2$ with a vertex in each of the four corners, where $B^2$ is the
total number of particles ($25$ in the present case) and for each box
we have computed the time history of the number of particles inside
the box, under the condition that at time $t=0$ all the particles are
in the leftmost bottom box.  The value for the lowest curve
(representing one given corner) is exactly the value of the number of
particles in the corner, while the other 3 curves have an offset of,
respectively, $30$, $60$ and $90$ for the 3 different corners.  The
transitions are always abrupt, and the particles spend the quasi
totality of their time confined in one corner. As one expects on
theoretical grounds the typical time scale on which a transition
occurs is much shorter than the time one has to wait to see the
transition. The system is waiting for a {\it critical fluctuation}
making a lot of unsuccessful attempts (the small spikes towards the
bottom of the figure).

\begin{figure}
\centering
\includegraphics[width=0.7\textwidth,angle=0]{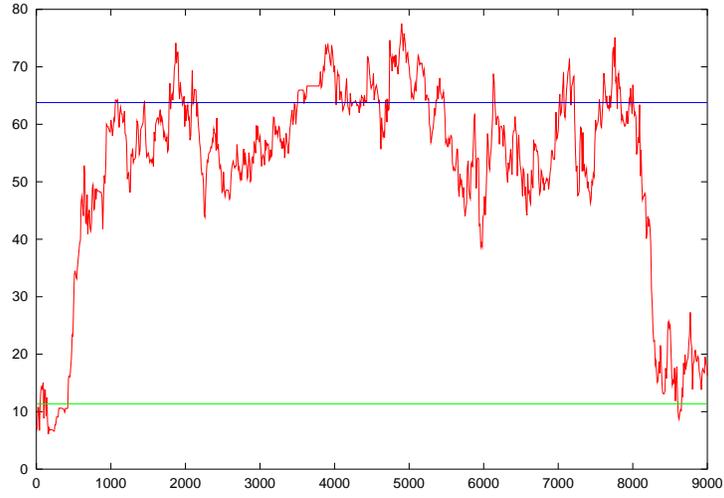}
\caption[a]{
Variance of the center of mass as a function of the time.
\protect\label{FIG-3}}
\end{figure}

In figure (\ref{FIG-3}) instead we have computed the ``variance'' of
the center of mass of the particles. We define

\begin{equation}
  \sigma^2(t) \equiv {1 \over B^2} \sum_i
  \left( x_i(t)-x^{(cm)}(t)\right)^2\ ,
\label{eq-cm}
\end{equation}
where $x_i(t)$ are, as usual, the positions of the particles on the
two dimensional square lattice, and $x^{(cm)}(t)$ is the center of
mass of the particles at time $t$.  We compare it to the two extreme
situations, where the particles form a compact blob (the lowest
straight line), and where the particles are divided in two, equal
sized, compact blobs in two adjacent corners (see the previous
discussion in \S 3.1). The time scale of this figure (in arbitrary
units again) is much smaller than the one of the previous figure, and
basically gives the time of a single transition. In other words we
have analyzed the time behavior of the above variance precisely during
the short (in the time scale of figure (\ref{FIG-2})) time interval in
which the transition takes place.  The outcome is that the particles
go from a compact blob (in a corner) to a compact blob (in a different
corner) passing through a situation where they are basically divided
in two groups of similar size.

\begin{figure}
\centering
\includegraphics[width=0.7\textwidth,angle=0]{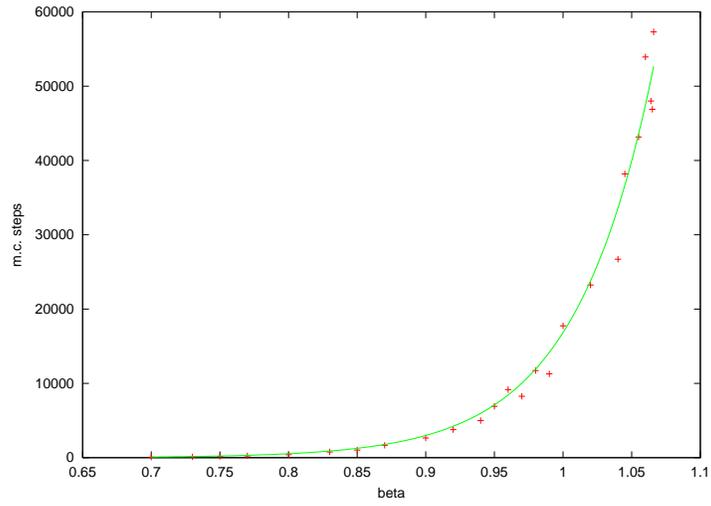}
\caption[a]{
Transition time as a function of $\beta$.
\protect\label{FIG-4}}
\end{figure}

The second part of our analysis considers scaling behaviors of the
(average) transition times $T(\beta,L)$. We look separately at the
dependence of $T$ over $\beta$ and over the size $L$.  In figure
(\ref{FIG-4}) we look at the $\beta$ dependence over the range
$[0.65,1.1]$ for a fixed side $L=20$ (and $25$ particles). Our best
fit to the form

\begin{equation}
  T(\beta) = A\ e^{\beta\delta H}
\end{equation}
is very good and gives $\Delta H \simeq 17.8$ (and $A\simeq 0.001$). 

\begin{figure}
\centering
\includegraphics[width=0.4\textwidth,angle=0]{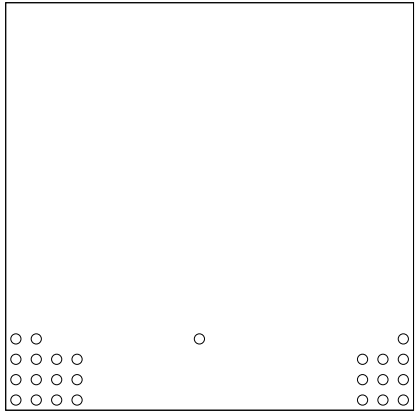}
\caption[a]{The dynamical mechanism we call {\em evaporation}.
See the text for further details.
\protect\label{FIG-5}}
\end{figure}

This is in very good agreement with the heuristic calculation of the
energy barrier assuming that the transition occur via the mechanism we
have called {\it evaporation}, that suggests that $\Delta H$ is of
order $20$.  We illustrate the mechanism in figure (\ref{FIG-5}). We
start with $25$ particles packed in the left down-most corner: here the
surface includes $10$ broken links. Now in the intermediate situation
of figure (\ref{FIG-5}), that is a typical intermediate particle
configuration, we have $19$ broken links. Clearly it would be less
expensive for the lonely particle to travel along the side of the
lattice, but the configuration we show has a higher entropy and we
observe these kind of processes with high frequency.  Since $\Delta H$
is equal to twice the difference in the number of broken links, in
this case $\Delta H = 18$. This is only an estimate, but gives the
correct order of magnitude. Notice that $\Delta H = 18$ is quite far
from the infinite volume saddle we have discussed before, that would
give $\Delta H = 4 (\sqrt{2}-1) B \simeq 8$: this is due to finite
size effects (both $B$ and $L$ are finite). We have verified that when
we change the number of particles (for example from $25$ to $36$) or
the definition of a {\it transition} from corner to corner we continue
to find the expected scaling behavior.

\section{Conclusions}

In this paper we have investigated, mainly numerically, the Kawasaki
dynamics ( a particular conservative particle--hole exchange) for the
low temperature ($\beta \approx 2\beta_c$) Ising lattice gas in a box
with free boundary conditions and side $L \approx 30$.  The number of
particles $N$ has been chosen in a such a way that phase segregation
occurs (typically $N\approx 25$). With the above choice of the
thermodynamic parameters the typical equilibrium configurations
consist of a (small compared to the total volume) square--like droplet
of particles which sits in one of the four corners because of the
chosen boundary conditions and a rarefied gas elsewhere.  Under the
Kawasaki dynamics the droplet of particles make rare and mostly
unsuccessful attempts to migrate to one of the empty corners. Based on
energetic considerations alone, thus neglecting entropic
contributions, we have envisaged two main best possible mechanisms for
the migration to take place, that we have named {\it sliding} (the
bubble of particles gets deformed into a rectangle, slides along one
side of the box until it reaches the opposite corner and finally
reconstructs the optimal square--like shape) and {\it evaporation}
(the particles in the bubble evaporate into the rarefied gas and
recollect together to a form a new bubble in another corner)
respectively. We have then performed intensive numerical investigation
to check whether sliding or evaporation is the prefered mechanism and
the scaling behavior with the inverse temperature $\beta$ of the
(average) transition time. Under various different definitions of the
bubble transition our results consistently indicate that {\it
evaporation} is the dominant effect and that the scaling law for the
average transition time $T(\beta)$ is of the form

\begin{equation}
  T(\beta) = A\ e^{\beta\delta H}\ ,
\end{equation}
where the energy barrier $\delta H$ is in very good agreement with
simple energetic considerations.

\end{document}